\documentclass[11pt,twoside]{book}
\usepackage{konkolyproc2}
\usepackage{longtable}
\usepackage{amsmath,amssymb}
\usepackage{graphicx}
\usepackage{lscape}
\usepackage{index}
\usepackage{natbib}
\usepackage{bigdelim}
\usepackage{multirow}
\makeindex

\begin{document}

\pagestyle{myheadings}
\setcounter{equation}{0}\setcounter{figure}{0}\setcounter{footnote}{0}
\setcounter{section}{0}\setcounter{table}{0}\setcounter{page}{1}
\markboth{Juh\'asz, Moln\'ar, \& Plachy}{Study of the RR Lyrae stars observed in 
K2 Campaign 3}
\title{A preliminary study of the RR Lyrae stars observed in \textit{K2} 
Campaign 3}
\author{\'Aron Juh\'asz$^1$, L\'aszl\'o Moln\'ar$^2$, \& Emese Plachy$^2$}
\affil{$^1$E\"otv\"os University, Budapest, Hungary\\
$^2$Konkoly Observatory, Research Centre for Astronomy and Earth Sciences, 
Hungarian Academy of Sciences, H-1121, Budapest, Konkoly Thege Mikl\'os \'ut 15-17, Hungary}

\begin{abstract}
We have started a comprehensive analysis of the \textit{Kepler} \textit{K2} Field 3 data set. 
Our goals are to assess the statistics of the sample, and to search
for peculiar stars. We found a candidate triple-mode RRab star, where
the first and ninth overtones also seem to be excited.
\end{abstract}

\section{The data sets of \textit{K2} Field 3}
The \textit{Kepler~K2} mission observed 79 fundamental and 1 first overtone RR~Lyrae 
variables in Field 3 for 69.2 days. Our goals are to investigate the Blazhko 
periods and examine the properties and occurrences of the various additional modes. 

The large number of stars in this sample gives us the opportunity to describe 
the abundance of each pulsation mode. To do this we need to investigate the 
reliability of the Field 3 data sets, especially the light curves of faint 
stars and stars close to the edge of the field of view. We summarize the 
first results in this paper.

At the release of \textit{K2} Field 3 data, NASA published two kinds of 
automatically generated light curves for long cadence targets. The first 
type was created with single aperture photometry. The second type used the 
same aperture, but the light curves underwent an optimizing and noise 
reducing method (PDCSAP). In the first part of the investigation we used 
these latter data.

Within Field 3 there are many PDCSAP light curves which frequently 
contain large and sudden jumps. The most determinative parameter is 
the distance of the star from the center of the field of view.   
The movement of the star's photocenter position is only 0.2-0.3 
pixel near the center, but this could be more than 2 pixels at the edge. Therefore 
in many cases the simple pixel photometry was impractical, because the PSF can 
move out from the mask or another star could contaminate it (especially in 
dense fields). This was the reason why we did our own photometry on the stars 
in many cases, occasionally using more than one aperture. Because of the large 
number of stars in this sample we hoped for finding some peculiar targets as well. 
Here we present one such case.

\newpage

\section{The possible FM-O1-O9 triple mode star EPIC 206280713}
The Fourier spectrum of the PDCSAP light curve shows one peak slightly above 
$1.5 f_0$. The original light curve does not show Blazhko modulation. 
We created our own tailor-made pixel photometry for the star with PyKE 
\citep{Still&Barclay2012}, which suggests that this star is in fact a Blazhko 
star with a low-amplitude and long-period modulation.

The frequency spectrum of the light curve confirms the peak at $1.5 f_0$ 
and clearly shows peaks on $0.5 f_0$ and $2.5 f_0$, the signs of period doubling. 
The most important discovery is the low-amplitude first overtone signal 
($f_0/f_1=0.7370$; $P_0=0.48146$~day, $P_1=0.35486$~day) and its linear 
combinations with the fundamental mode. The frequency spectrum of EPIC 206280713 
closely resembles that of RR Lyr itself, and provides another opportunity for 
comparison with triple-mode non-linear hydrodynamic model calculations 
\citep{Molnar2012}.
These periods put the star in the same region in the Petersen diagram where 
the modulated double-mode stars are also located \citep{jurcsik2015,smolec2015}.

\begin{figure}[!ht]
\includegraphics[width=1.0\textwidth]{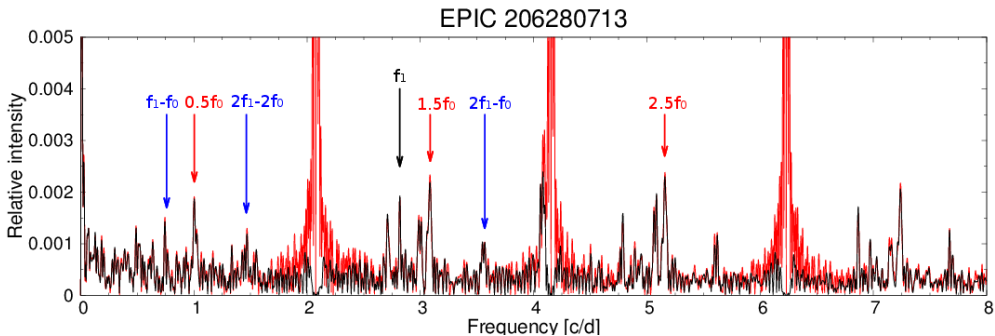}
\caption{Fourier-spectrum of the tailor-made photometry. Black arrow: the 
possible first overtone. Red arrows: peaks at $0.5 f_0$, $1.5 f_0$, $2.5 f_0$. 
Blue arrows: the linear combinations of $f_0$ and $f_1$.} 
\label{juhasz-fig1} 
\end{figure}

\section*{Acknowledgements}
This work has used \textit{K2} data selected and proposed by the RR Lyrae and Cepheid 
Working Group of the Kepler Asteroseismic Science Consortium (proposal GO3040). 
This project has been supported by the Lend\"ulet-2009 and LP2014-17 Programs 
of the Hungarian Academy of Sciences, and by the NKFIH K-115709 and PD-116175 
grants of the Hungarian National Research, Development and Innovation Office.


\begin{thebibliography}{}      
\bibitem[Jurcsik et~al.(2015)]{jurcsik2015} 
Jurcsik, J., Smitola, P., Hajdu, G., et al. 2015, ApJS, 219:25
\bibitem[Moln\'ar et~al.(2012)]{Molnar2012}
Moln\'ar, L., Koll\'ath, Z., Szab\'o, R., et al. 2012, ApJ, 757:L13
\bibitem[Smolec et~al.(2015)]{smolec2015} 
Smolec, R., Soszy\'nski, I., Udalski, A., et al. 2015, MNRAS, 447, 3756
\bibitem[Still \& Barclay(2012)]{Still&Barclay2012} 
Still, M., Barclay, T. 2012, Astrophysics Source Code Library, ascl:1208.004
\end{thebibliography}
\end{document}